\documentclass[aps,prl,twocolumn,superscriptaddress,showpacs]{revtex4}
\usepackage{graphicx}
\usepackage{amssymb}
\usepackage{amsmath}
\usepackage{rotating,dcolumn}
\begin{document}

\title{Parity Broken Chiral Spin Dynamics in Ba$_3$NbFe$_3$Si$_2$O$_{14}$}

\author{M. Loire}
\affiliation{Institut N\'eel, CNRS \& Universit\'e Joseph Fourier,
BP166, 38042 Grenoble Cedex 9, France}
\author{V. Simonet}
\affiliation{Institut N\'eel, CNRS \& Universit\'e Joseph Fourier,
BP166, 38042 Grenoble Cedex 9, France}
\author{S. Petit}
\affiliation{Laboratoire L\'eon Brillouin, CEA-CNRS, CE-Saclay, F-91191 Gif sur Yvette, France}
\author{K. Marty}
\affiliation{Institut N\'eel, CNRS \& Universit\'e Joseph Fourier,
BP166, 38042 Grenoble Cedex 9, France} \affiliation{Oak Ridge National Laboratory, Oak Ridge, Tennessee 37831 USA }
\author{P. Bordet}
\affiliation{Institut N\'eel, CNRS \& Universit\'e Joseph Fourier,
BP166, 38042 Grenoble Cedex 9, France}
\author{P. Lejay}
\affiliation{Institut N\'eel, CNRS \& Universit\'e Joseph Fourier,
BP166, 38042 Grenoble Cedex 9, France}
\author{J. Ollivier}
\affiliation{Institut Laue Langevin,  BP156, 38042 Grenoble
Cedex 9, France}
\author{M. Enderle}
\affiliation{Institut Laue Langevin,  BP156, 38042 Grenoble
Cedex 9, France}
\author{P. Steffens}
\affiliation{Institut Laue Langevin,  BP156, 38042 Grenoble
Cedex 9, France}
\author{E. Ressouche}
\affiliation{Institut de
Nanosciences et Cryog\'enie, SPSMS/MDN, CEA-Grenoble, 38054
Grenoble Cedex 9, France}
\author{A. Zorko}
\affiliation{"Jo\v{z}ef Stefan" Institute, Jamova 39, 1000
Ljubljana, Slovenia}
\affiliation{EN $\rightarrow$ FIST Centre of Excellence, Dunajska 156, SI-1000 Ljubljana, Slovenia}
\author{R. Ballou}
\email[]{rafik.ballou@grenoble.cnrs.fr} \affiliation{Institut
N\'eel, CNRS \& Universit\'e Joseph Fourier, BP166, 38042 Grenoble
Cedex 9, France}

\date{\today}

\begin{abstract}
The spin wave excitations emerging from the chiral helically modulated 120$^{\circ}$ magnetic order in a langasite Ba$_3$NbFe$_3$Si$_2$O$_{14}$ enantiopure crystal were investigated by unpolarized and polarized inelastic neutron scattering. A dynamical fingerprint of the chiral ground state is obtained, singularized by (i) spectral weight asymmetries answerable to the structural chirality and (ii) a full chirality of the spin correlations observed over the whole energy spectrum. The intrinsic chiral nature of the spin waves elementary excitations is shown in absence of macroscopic time reversal symmetry breaking.

\end{abstract}


\maketitle

\begin{figure}[t]
\includegraphics[bb=20 350 642 575,scale=0.45]{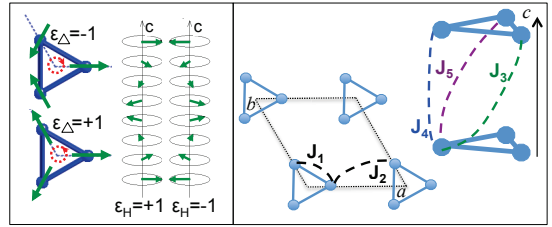}
\caption{from left to right:  the two triangular chiralities $\epsilon_{\triangle}=\pm1$, the two helicities $\epsilon_H=\pm1$, 
and the magnetic exchange paths $J_1$ to $J_5$ in Ba$_3$NbFe$_3$Si$_2$O$_{14}$.}
\label{f.0}
\end{figure}

Ubiquitous in nature, chirality is what distinguishes a phenomenon from its materialization in a mirror \cite{Wagniere2007}. Condensed matter physics and especially magnetism provide a rich playground to investigate this phenomenon as it appears naturally to describe the arrangement of magnetic moments in non collinear magnets. The chirality accounts there for the sense of rotation of the spins on moving along an oriented line. The latter can be a straight line around which the spins form an helix, or an oriented loop, for instance a triangle, at the summits of which the spins are orientated at 120$^{\circ}$ of each other (see Fig. \ref{f.0}). For such coplanar spin configurations, it can be uniquely characterized on a bond connecting two consecutive spins $\vec S_i$ and $\vec S_j$ by the parity breaking vector product $\vec{\chi}_{ij} = \vec S_i \wedge \vec S_j$, defining the magnetic vector chirality. As far as domains of opposite chiralities are equipopulated, the chirality is usually averaged to zero in centrosymmetric magnets. However, neutron scattering experiments evidenced a pure macroscopic chiral phase in non-centrosymmetric compounds \cite{Ishida1985,Janosheck2010}, as for instance the Ba$_3$NbFe$_3$Si$_2$O$_{14}$ langasite \cite{Marty2008b}. 

While these studies focused on the ground state properties, the way the associated excited states inherit from the chiral properties remains an open issue. In this letter, we address this intriguing question, investigating the spin dynamics in a Ba$_3$NbFe$_3$Si$_2$O$_{14}$ single crystal by means of polarized neutron scattering experiments. Our study provides evidence for a fully chiral spin excitation spectrum over the whole energy range. Chiral dynamics was already evidenced in a variety of magnetic materials \cite{Moon1969,Lovesey1998,Gukasov1999,Roessli2002,Grohol2005,Braun2005,Plakhty2006}, but we emphasize that these studies were carried out under a magnetic field, revealing the chirality through a macroscopically breaking of the time reversal symmetry. In the Ba$_3$NbFe$_3$Si$_2$O$_{14}$ case, the chiral dynamics is observed in zero field, thus ascribable solely to the space inversion symmetry breaking. To our knowledge, this has never been reported so far. 

The langasite Ba$_3$NbFe$_3$Si$_2$O$_{14}$ presents static chirality properties \cite{Marty2008b,Marty2008a,Marty2010} hereafter briefly recalled. It crystallizes in the acentric space group P321 which confers the compound a {\bf structural chirality} denoted $\epsilon_T=+(-)1$ for right(left)-handedness. The magnetic carriers are the Fe$^{3+}$ ions ($S=5/2$, $L=0$), arranged on small triangle units (trimers) whose centers form triangular lattices perpendicular to the trigonal $c$ axis (see Fig. \ref{f.0}). A magnetic order sets in below the N\'eel temperature $T_N=27~K$ with spins lying in the ($a$, $b$) plane and with the same 120$^{\circ}$ configuration on each trimer. This arrangement is helically modulated along the $c$ axis with the period $1/\tau \approx 7$ (see Fig. 4 of ref. \onlinecite{Marty2008b}). This magnetic order exhibits a {\bf single chirality of the helical order} (named helicity), denoted $\epsilon_H$, and a {\bf single triangular chirality}, denoted $\epsilon_{\triangle}$ (see Fig. \ref{f.0}). We note that $\vec{\chi}_{ij} = S^2\sin(2\pi\tau)\epsilon_H \hat c$ for consecutive spins along $c$ and $\sum_{\Delta \circlearrowleft}\vec{\chi}_{ij} = (3\sqrt 3/2)S^2\epsilon_{\triangle}\hat c$ for the spins of an oriented triangle, with $\epsilon_H$ and $\epsilon_{\triangle}$ equal to $+1(-1)$ for the right(left)-handed sense of rotation.

\begin{figure}[t]
\includegraphics[bb=20 80 842 790,scale=0.44]{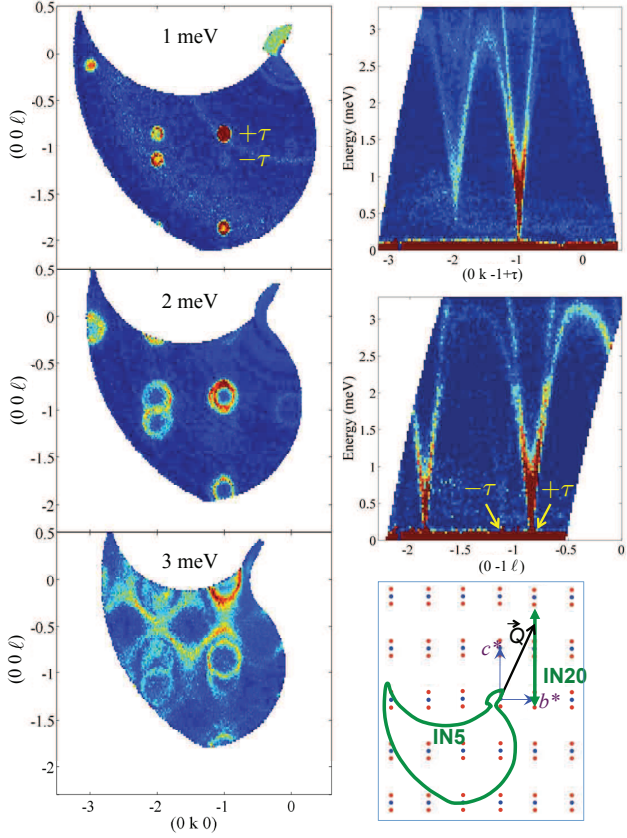}
\caption{Inelastic neutron scattering intensities at 1.6~K in the
($b^*$, $c^*$) scattering plane on the time-of-flight spectrometer IN5. Displayed are intensity maps at constant transfer energy (energy-cuts) and intensity maps along $(0,k,-1+\tau)$ and $(0,-1,\ell)$ lines in the reciprocal space ($\vec{Q}$-cuts). In the bottom right sketch, the green area and arrow show the probed reciprocal vectors on the IN5 and IN20 spectrometers.}
\label{f.1}
\end{figure}

\begin{figure}[t]
\includegraphics[bb=50 120 595 745,scale=0.5]{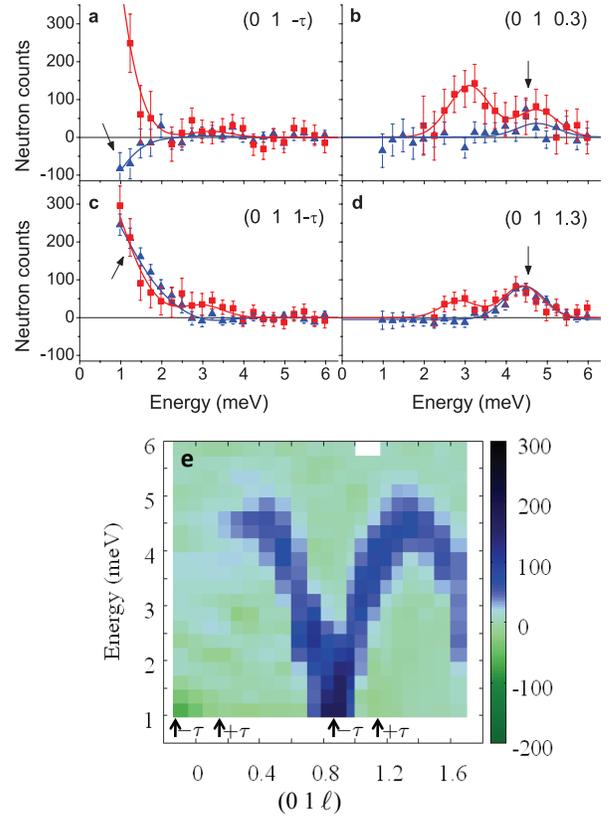}
\caption{(a-d) Magnetic scattering $S(\vec{Q},\omega)$ (red) and chiral contribution $C(\vec{Q},\omega)$ (blue) as a function of the transfer energy at selected scattering wavevectors $(0, 1, \ell)$, extracted at 1.5 K by longitudinal neutron polarimetry on the triple-axis spectrometer IN20.  The spectral weight is only significant for the branch emerging from the $-\tau$ magnetic satellites. (e)  Dispersion of the chiral scattering $C(\vec{Q},\omega)$ obtained by gathering the measurements at all $\ell$.} \label{f.3}
\end{figure}

\begin{figure}[t]
\includegraphics[bb=40 120 842 790,scale=0.46]{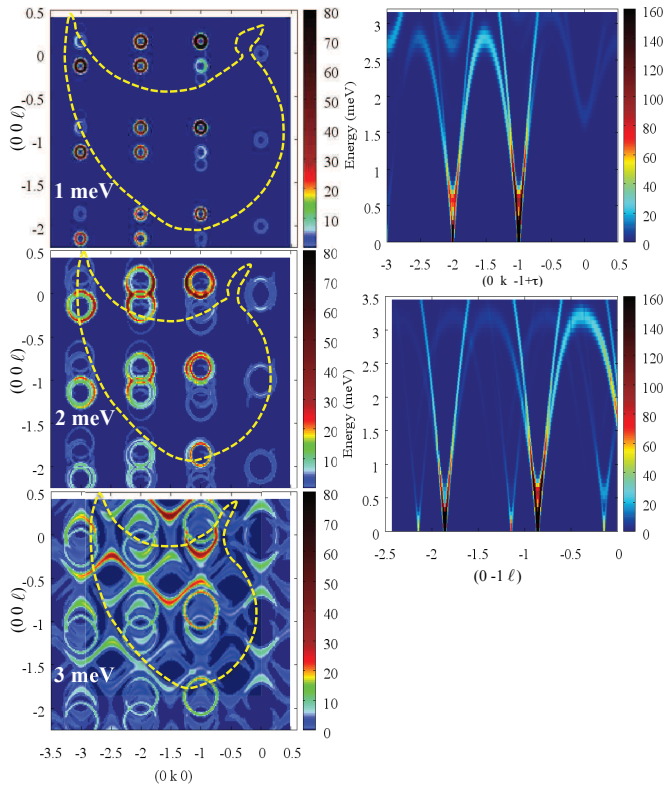}
\caption{Dynamical structure factor calculated in the linear
approximation from the minimal model (see
text) to compare with the measurements of Fig. \ref{f.1}.}
\label{f.2}
\end{figure}

According to \cite{Blume, Maleyev}, polarized inelastic neutron scattering can probe the dynamical spin chirality. This extends the concept of static chirality to the dynamics by following the orientation changes during coherent motions of spins in space and time. Its statistical average is proportional to the antisymmetric off-diagonal dynamical magnetic susceptibility \cite{Maleyev95,Syromyatnikov}. We thus conducted experiments on the triple-axes spectrometer IN20 at the ILL in its polarized neutron set up, with polarizing Heusler crystals as monochromator and analyzer. We used the CRYOPAD device, to obtain strict zero-field environment at the sample position, and to prepare incoming and outgoing neutron polarization independently. $E_f =$14.7 meV was kept fixed yielding an energy resolution of $\simeq$ 1 meV and second order contaminations were removed by a PG(002) filter. The crystal, aligned with $(0, k, \ell)$ as scattering plane, was cooled down to 1.5 K in an Orange cryostat. Different cross sections were measured, $\sigma_{\uparrow\uparrow}$, $\sigma_{\uparrow\downarrow}$ and $\sigma_{\downarrow\uparrow}$ where the subscripts indicate the incoming/outgoing neutron polarization parallel $\uparrow$ or antiparallel $\downarrow$ to the scattering vector $\vec{Q}$. Additional cross sections were measured at selected points in $\vec{Q}$ and energy transfers $\hbar\omega$, allowing to separate the magnetic signal from background, and to verify that phonon and incoherent scattering are negligible in the investigated region. $\sigma_{\uparrow\uparrow}$ could therefore be used as background above $\approx$1.5 meV. The magnetic dynamical structure factor was then obtained by $S(\vec{Q},\omega) = 
{1\over2\pi\hbar}\int \langle M_y(\vec{Q},0)M_y(-\vec{Q},t) + M_z(\vec{Q},0)M_z(-\vec{Q},t)\rangle e^{-i\omega t}dt$ = $(\sigma_{\uparrow\downarrow}+\sigma_{\downarrow\uparrow})/2-\sigma_{\uparrow\uparrow}$ and the chiral magnetic dynamical scattering by $C(\vec{Q},\omega) = {1\over2\pi\hbar}\int \langle M_y(\vec{Q},0)M_z(-\vec{Q},t) - M_z(\vec{Q},0)M_y(-\vec{Q},t)\rangle e^{-i\omega t}dt$ = $(\sigma_{\uparrow\downarrow}-\sigma_{\downarrow\uparrow})/2$. Here, $M_{y,z}(\vec{Q},t)$ are the Fourier transformed spin components at time $t$ perpendicular to $\vec Q$ within ($y$) or perpendicular ($z$) to the scattering plane.  

To get first a global overview of the spin waves, an experiment using unpolarized neutrons was performed on the time-of-flight spectrometer IN5 at the ILL on a single crystal in rotation around the vertical zone axis $a$ \cite{IN5_NN}. The incident wavelength was fixed to 4 {\AA}, the chopper speed to 16000 rpm, yielding an elastic energy resolution $\simeq$ 0.1 meV. Standard corrections and a background (high temperature scan) subtraction were performed. The data were then reduced with the Horace suite software \cite{horace_ISIS} to obtain the excitation spectra as function of ($\vec{Q}$,$\omega$). 

Fig. \ref{f.1} summarizes our inelastic unpolarized neutron scattering results obtained on IN5. Two spin wave branches emerging from the $\pm\tau$ magnetic satellites are identified \cite{footnote3}. They form delicate arches, with different maximum energies (lower branch clearly visible on IN5, maximum of the upper branch only observable on IN20). One of the branches is gapped with a minimum at around 0.4~meV whereas the other branch is found ungapped down to the resolution (0.1~meV). A first unusual observation is the difference of intensities of the excitations emerging from the $\pm\tau$ satellites associated to a node of the reciprocal lattice. This effect strongly depends on the considered reciprocal lattice node and is clearly visible for instance around (0,-1,-1) in Fig. \ref{f.1}. This is a signature of the structural chirality as detailed below. 

Fig. \ref{f.3} gathers our inelastic polarized neutrons scattering results of IN20. Energy-scans at constant $\vec Q$ were performed for different $\ell$ values along $(0, 1, \ell)$ with $\ell$ ranging from -$\tau=-{1\over 7}$ to 1.7 (see sketch in Fig. \ref{f.1}). The magnetic scattering $S(\vec{Q},\omega)$ (red 
lines in Figs. \ref{f.3}a-d) confirms the magnetic origin of the two modes, although they are not as well separated as on IN5, due to the lower energy resolution. By extracting the chiral contribution $C(\vec{Q},\omega)$ (in blue), it is found that the lower mode is achiral ($C(\vec{Q},\omega)$=0) whereas the upper mode has finite chirality (see Fig. \ref{f.3}d and e). In addition, $C(\vec{Q},\omega)$ is positive and has a tendency to become negative for negative $\ell$ (compare Fig. \ref{f.3}a and c at low energy). This change of sign is due to the fact that the neutron probes the moment component perpendicular to $\vec Q$, $C(\vec{Q},\omega)$ corresponding then to the projection of the Fourier transformed dynamical chirality onto $\vec Q$. This result actually perfectly reflects a globally unchanged chirality of the upper branch all the way up to the maximum of the dispersion. At large $\ell$ values, $C(\vec{Q},\omega)$ equals the amplitude of $S(\vec{Q},\omega)$ (see Fig. \ref{f.3}d), pointing at a full chirality of the upper branch without chirality mixing. 

\begin{figure}[t]
\includegraphics[bb=20 140 842 780,scale=0.48]{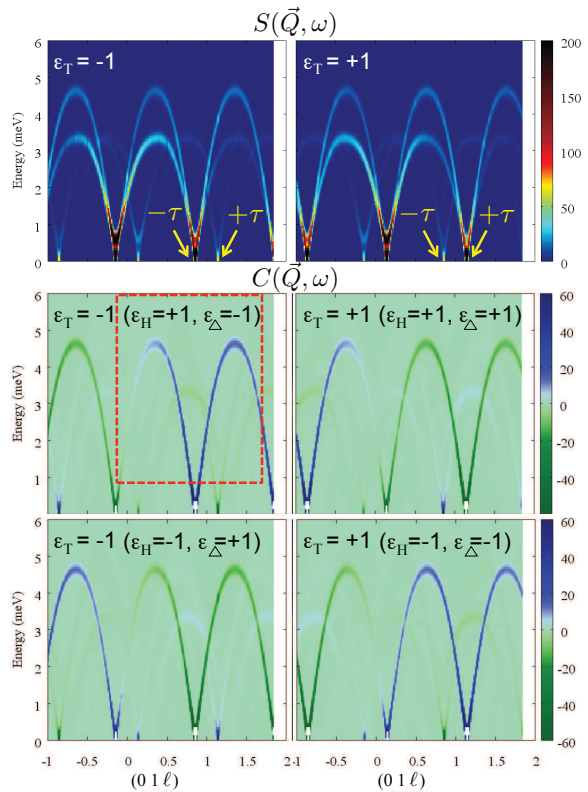}
\caption{Top: computed magnetic scattering $S(\vec{Q},\omega)$ for the two structural chiralities $\epsilon_T\pm1$. Below each of them, computed chiral contribution $C(\vec{Q},\omega)$ for the 2 associated magnetic ground states $(\epsilon_H,\epsilon_{\triangle}=\epsilon_H\epsilon_T)$. The red frame points out the calculation agreeing with the experiment in Fig. \ref{f.3}.e.} \label{f.4}
\end{figure}

To shed light on this remarkable result, a characterization of the ground state can be done from a careful analysis of the spin waves. We propose a model, based on mean-field calculations, that accounts qualitatively for the magnetic order \cite{Marty2008b} (see Fig. \ref{f.0}). The in plane 120$^\circ$ spin arrangement is first stabilized by antiferromagnetic intra-trimer $J_1$ and inter-trimer $J_2$ interactions within the ($a$, $b$) planes. Next, three inter-plane interactions $J_3$ to $J_5$, connect each moment to the $3$ moments of the upper/lower trimer along the $c$ axis. The acentric structure imposes different $J_3$ and $J_5$, oppositely twisted around $c$. The dominant one drives the helical modulation according to the structural chirality ($J_5>J_3$,$J_4$ for $\epsilon_T=-1$ and $J_3>J_4$,$J_5$ for $\epsilon_T=+1$). The two weakest interactions allow tuning the periodicity $\tau$ of the helix. This model implies that the magnetic and structural chiralities are related ($\epsilon_{\triangle}=\epsilon_T\epsilon_H$): for the investigated crystal, $\epsilon_T=-1$ is observed (strong $J_5$), imposing opposite senses of rotation for the helicity and the triangular chirality. We are left with two different solutions (out of four) for the magnetic chiralities ($\epsilon_H=+1$, $\epsilon_{\triangle}=-1$) and ($\epsilon_H=-1$, $\epsilon_{\triangle}=+1$). The observed ultimate selection of a single magnetic chirality was proposed \cite{Marty2008b} to originate from the antisymmetric Dzyaloshinskii-Moriya (DM) exchange interaction $\vec D. (\vec S_i\times\vec S_j)$ with $\vec D$ the DM vector, allowed in absence of inversion symmetry center between spins $\vec S_i$ and $\vec S_j$. It suffices to consider the DM interaction inside the trimer with the same DM vector along the $c$ axis for the three bonds \cite{Moriya1960}. This favors planar spin components and its sign selects a triangular chirality $\epsilon_{\triangle}$, and hence a helicity $\epsilon_H$ since $\epsilon_{\triangle}=\epsilon_T\epsilon_H$.

Using the standard Holstein-Primakov formalism in the linear approximation \cite{Holstein}, we computed $S(\vec{Q},\omega)$ and $C(\vec{Q},\omega)$ at zero temperature. As shown by the comparison of Fig. \ref{f.1} and \ref{f.2}, a good agreement between experiment and calculation is achieved with the exchange parameters (in meV) $J_1=0.85\pm 0.1$, $J_2=0.24 \pm 0.05$, $J_3=0.053 \pm 0.03$, $J_4=0.017$ and $J_5=0.24 \pm 0.05$ and a DM vector along c of $\approx1\%|J_1|$. The latter was checked to produce the lower branch gap and to select a triangular chirality \cite{footnote}. $J_3, J_4, J_5$ were constrained to fulfill the $\tau=1/7$ conditions and the set of best parameters yields a Curie-Weiss temperature of 191~K, within less than 10\% of the values obtained from  susceptibility measurements \cite{Marty2008b,Zhou}. Other models have been proposed from spin wave measurements in Ba$_3$NbFe$_3$Si$_2$O$_{14}$, 
but they fail in reproducing all the features revealed in our measurements \cite{Zhou2,Stock}. Interestingly, as shown by the 
calculations in Fig. \ref{f.4}, $S(\vec{Q}, \omega)$ reflects the structural chirality: the asymmetric spectral weight of the branches emerging from the $+\tau$ and $-\tau$ satellites is inverted for $\epsilon_T=\pm1$. On the other hand, for a given $\epsilon_T$, $S(\vec{Q}, \omega)$ does not depend on $\epsilon_{\Delta}$. The two corresponding magnetic ground states yield however $C(\vec{Q}, \omega)$ with opposite sign (see Fig. \ref{f.4}). $S(\vec{Q}, \omega)$ and $C(\vec{Q}, \omega)$ therefore provide with a strong dynamical fingerprint of the ground state chirality. Finally, the calculated chiral scattering for ($\epsilon_H=+1$, $\epsilon_{\triangle}=-1$) reproduces very well the measurements of Fig. \ref{f.3}: its sign, the absence of chirality of the lower branch and the full chirality of the upper branch. The calculated correlation functions show that the spins components involved are along $a$ and $b$ for the upper branch, and only along $c$ for the lower branch, which yields a zero spin cross product and explains the absence of chirality. 

In summary, the spin waves excitation modes emerging from the totally chiral magnetic order in a structurally enantiopure single
crystal of Ba$_3$NbFe$_3$Si$_2$O$_{14}$ are extremely unusual.  A spectral asymmetry is evidenced by inelastic unpolarized neutron scattering as a fingerprint of the crystal chirality.  In addition, the use of polarized neutrons and polarization analysis led to the discovery that the spin waves with in-plane spin 
correlations are fully chiral at all energies. Chiral dynamics was predicted from spin waves calculation for a Heisenberg triangular antiferromagnet with unbalanced domains of opposite triangular chirality \cite{Syromyatnikov}. This condition is obviously realized with a single chirality domain in Ba$_3$NbFe$_3$Si$_2$O$_{14}$. However, the absence of chirality mixing of the spin dynamics is not only observed close to the magnetic satellites as calculated \cite{Syromyatnikov} but up to the maximum of the dispersion branches ($\approx$4.8 meV). It finally will be emphasized that the crystal was in a strictly zero magnetic field (even the Earth's magnetic field is excluded from it) during the experiment on IN20. This establishes the observation of a dynamical chirality unbiased by macroscopic time-reversal symmetry breaking, but solely associated with the space inversion symmetry breaking.

\acknowledgments This work was financially supported by the ANR
06-BLAN-01871. We would like to thank B. Canals and A. Ralko for
fruitful discussions, A. Hadj-Azzem, J. Balay and J. Debray for the sample preparation.


\begin{thebibliography}:

\bibitem{Wagniere2007} G. H. Wagni\`ere, {\it On Chirality and the Universal Asymmetry}, WileyÐVCH, Z\"urich, Weinheim, 2007.
\bibitem{Ishida1985} M. Ishida, {\it et al.}, J. Phys. Soc. Jap. {\bf 54}, 2975 (1985).
\bibitem{Janosheck2010} M. Janoscheck, {\it et al.}, Phys. Rev. B {\bf 81}, 094429 (2010).
\bibitem{Marty2008b} K. Marty, {\it et al.}, Phys. Rev. Lett. {\bf 101}, 247201 (2008).
\bibitem{Roessli2002} B. Roessli, {\it et al.},  Phys. Rev. Lett. {\bf 88}, 237204 (2002).
\bibitem{Grohol2005} D. Grohol, {\it et. al.} Nature Materials {\bf 4}, 323 (2005).
\bibitem{Plakhty2006} V. P. Plakhty, {\it et. al.}, Physica B {\bf 385}, 288 (2006).
\bibitem{Braun2005} H. Braun,  {\it et. al.}, Nature Physics,  {\bf 1}, 159 (2005).
\bibitem{Moon1969}  R. M. Moon, T. Riste and W.C. Koehler, Phys. Rev. {\bf 181}, 920  (1969).
\bibitem{Lovesey1998} S. W. Lovesey and G. I. Watson, J. Phys.: Condens. Matter {\bf 10}, 6761 (1998)
\bibitem{Gukasov1999} A. Gukasov, Physica B {\bf 267}, 97 (1999).
\bibitem{Marty2008a} K. Marty, {\it et al.}, J. Magn. Magn. Mat. {\bf 321}, 1778 (2008).
\bibitem{Marty2010} K. Marty, {\it et al.}, Phys. Rev. B {\bf 81}, 054416 (2010).
\bibitem{Blume} M. Blume, Phys. Rev. {\bf 130}, 1670 (1963).
\bibitem{Maleyev} S. V. Maleyev, V. G. Bar'yakhtar, and R. A. Suris, Fiz. Tverd. Tela {\bf 4}, 3461 (1962).
\bibitem{Maleyev95} S. V. Maleyev, Phys. Rev. Lett {\bf 75}, 4682 (1995).
\bibitem{Syromyatnikov} A. V. Syromyatnikov, Phys. Rev. B {\bf 71}, 144408 (2005).
\bibitem{IN5_NN} J. Ollivier, {\it et al.}, Neutron News {\bf 21} (2010) 22.
\bibitem{horace_ISIS} T.G. Perring, {\it et al.}, (2009) unpublished, http://horace.isis.rl.ac.uk/.
\bibitem{footnote3} Less intense modes are also detected dispersing from the $\pm 2\tau$ satellites and from the nodes of the reciprocal lattice.
\bibitem{Moriya1960} T. Moriya, Phys. Rev. {\bf 120}, 91 (1960). 
\bibitem{Holstein} T. Holstein, and H. Primakoff, Phys. Rev. {\bf 58}, 1098 (1940).
\bibitem{footnote} A magnetocrystalline anisotropy and a symmetry-allowed component of the intra-triangle DM vector along the
bond connecting two spins were also considered without further improving the neutron data interpretation. 
\bibitem{Zhou} H. D. Zhou, {\it et al.}, Chem. Mater. {\bf 21}, 156 (2009).
\bibitem{Stock} C. Stock, {\it et al.} arXiv:1007.4216.
\bibitem{Zhou2} H. D. Zhou, {\it et al.}, Phys. rev. B. {\bf 82}, 132408 (2010).
\end{thebibliography}
\end{document}